\documentclass{scrartcl}
\usepackage[numbers]{natbib}
\usepackage{fontenc}
\usepackage{shadethm}
\usepackage{amsthm}
\usepackage{float}
\usepackage{stmaryrd}
\usepackage{mathrsfs} 
\usepackage{amsmath}
\usepackage{amssymb}
\usepackage{wasysym}
\usepackage{sectsty}
\usepackage{hyperref}
\usepackage[a4paper, left=2.7cm, right=2.7cm]{geometry}
\usepackage{chngcntr}
\counterwithin*{equation}{section}

\usepackage{cleveref}
\DeclareMathAlphabet{\mathpzc}{OT1}{pzc}{m}{it}

\sectionfont{\normalsize\centering}
\subsectionfont{\footnotesize\centering}
\newcommand{\norm}[1]{\left\lVert#1\right\rVert}

\theoremstyle{plain}
\newtheorem{thm}{Theorem}[section] % reset theorem numbering for each chapter

\theoremstyle{definition}
\newtheorem{defn}[thm]{Definition} % definition numbers are dependent on theorem numbers
\newtheorem{lem}[thm]{Lemma}
\newtheorem{prop}[thm]{Proposition}

\newtheorem{cor}[thm]{Corollary}

\def\XXint#1#2#3{{\setbox0=\hbox{$#1{#2#3}{\int}$ }
		\vcenter{\hbox{$#2#3$ }}\kern-.6\wd0}}

\usepackage[german,english]{babel}

\newcounter{MPequ}
\newenvironment{MPEquation}
{\stepcounter{MPequ}%
	\addtocounter{equation}{-1}%
	\equation}
{\endequation}

\begin{document}\selectlanguage{english}
\begin{center}
\normalsize \textbf{\textsf{Existence and characterisation of magnetic energy minimisers on oriented, compact Riemannian $3$-manifolds with boundary in arbitrary helicity classes}}
\end{center}
\begin{center}
	Wadim Gerner
\end{center}
{\small \textbf{Abstract:} In this paper we deal with the existence, regularity and Beltrami field property of magnetic energy minimisers under a helicity constraint. We in particular tackle the problem of characterising local as well as global minimisers of the given minimisation problem. Further we generalise Arnold's results concerning the problem of finding the minimum magnetic energy in an orbit of the group of volume-preserving diffeomorphisms to the setting of abstract manifolds with boundary.}
\section{Introduction}
Magnetohydrodynamics is concerned with the dynamics of electrically conducting fluids under the influence of an external electromagnetic field. Of particular interest is the special case of an ideal fluid, that is, a perfectly electrically conducting, incompressible, Newtonian fluid of constant viscosity. The dynamics in this case are governed by the equations of ideal magnetohydrodynamics (IMHD). Most notably in the ideal case is the fact that the electric and magnetic fields are perpendicular to one another, which gives rise to a conserved quantity, the so called \textit{helicity}. More precisely, if we consider a simply connected, bounded domain $\Omega\subset \mathbb{R}^3$ with smooth boundary and impose the boundary condition that the magnetic field $B$ is always tangent to the boundary, then one formally checks that the quantity $\mathcal{H}(B):=\int_\Omega A\cdot Bd^3x$, where $A$ is any vector potential of $B$, is in fact constant in time. The conservation of helicity was first observed by Woltjer \cite{W58} and a physical interpretation was, for instance, given by Moffatt \cite{M69}. The helicity may be regarded as a measure of linkage of distinct field lines of the underlying magnetic field. A similar interpretation of the helicity on closed $3$-manifolds with vanishing first and second de Rham cohomology groups was derived by Arnold \cite{A74} and Vogel \cite{V03}. In particular they prove that the helicity of a smooth divergence-free vector field coincides with the average linking number of the field lines of the considered vector field.
\newline
Helicity has been widely studied in mathematics and physics, see for example \cite{W58}, \cite{M69}, \cite{A74}, \cite{AL91} \cite{MR92}, \cite{AK98}, \cite{CDG00} to name a few. More recent works include \cite{CP10}, where the authors generalise the notion of helicity to higher dimensions and provide a characterisation of diffeomorphisms under which helicity is preserved, and \cite{EPT16}, where it is shown that helicity is essentially the only regular integral invariant of exact vector fields which is preserved under the action of volume-preserving diffeomorphisms.
\newline
To motivate our study of certain minimisation problems let us shortly recall what asymptotical behaviour of magnetohydrodynamical systems is expected. Following the exposition in \cite{A74} and \cite{AK98} one can argue in several steps. First of all one feature of IMHD is that magnetic field lines are frozen into the fluid. This is known as Alfv\'{e}n's theorem \cite{A42} and more precisely means that every fluid particle, lying on some initial magnetic field line continuous to lie on that same field line for all times. Since the dynamics of the fluid are governed by the Navier-Stokes equations, where the force-density is given by the Lorentz force, there is a coupling of the magnetic energy with the kinetic energy of the system, and one expects that magnetic energy is transformed into kinetic energy. If we now impose the no-slip condition, meaning that the velocity vector field vanishes on the boundary, one expects the total energy to decrease over time due to the dissipative nature of the equations involved. Overall the total energy should decrease until the fluid eventually comes to rest. Then Alfv\'{e}n's theorem implies that the magnetic field, being frozen-in, also comes to rest and becomes static. In addition the terminal magnetic field configuration must be a local minimiser of the magnetic energy, because otherwise the excess energy would be transformed into kinetic energy and yet again would be dissipated.
\newline
So local minimisers of the magnetic energy in fixed helicity classes are potential terminal static configurations for suitable initial magnetic field configurations in IMHD. This motivates the following minimisation problem
\begin{MPEquation}
	\label{MP1}
	\mathcal{E}(B):=\int_{\Omega}B^2d^3x\rightarrowtail \text{min,} \quad \mathcal{H}(B)=\int_{\Omega}B\cdot Ad^3x=\text{const.},
\end{MPEquation}
where $\mathcal{E}$ is the magnetic energy and $A$ is any vector potential of the divergence-free (static) magnetic field $B$. 
\newline
There is, however, yet another minimisation problem related to IMHD. We recall that the time evolution of the magnetic field $B$ is given by the equation
\[
\partial_t B=-[v,B],
\]
where $v$ is the corresponding velocity vector field of the fluid and $[\cdot,\cdot]$ denotes the Lie-bracket of vector fields. From this equation one can conclude that $B$ at time $t$ can be expressed as: $B=(\psi_t)_{*}B_0$, where $\psi_t$ denotes the flow of $v$, $(\psi_t)_{*}B_0$ denotes the pushforward induced by $\psi_t$, and $B_0$ is the initial configuration of $B$ at time $t=0$. Note that since IMHD is concerned with incompressible fluids, we have $\text{div}(v)=0$ and hence $\psi_t$ defines a volume-preserving diffeomorphism. Therefore we might as well look at the minimisation problem, where we minimise the magnetic energy $\mathcal{E}(B)$ on the class $\mathcal{V}_{B_0}(\Omega)$ consisting of the vector fields $B$ such that there is a volume-preserving diffeomorphism $\psi:\overline{\Omega}\rightarrow \overline{\Omega}$ with $B=\psi_{*}B_0$, for some fixed (initial) configuration $B_0$
\begin{MPEquation}
	\label{MP2}
	\mathcal{E}(B) \rightarrowtail \text{min},\quad B\in \mathcal{V}_{B_0}(\Omega).
\end{MPEquation}
We prove in \cref{T21} that solutions of the problem (\ref{MP1}) are Beltrami-fields, i.e., they are eigenfields of the curl operator corresponding to non-zero eigenvalues. In \cref{T23} we show that on the other hand solutions of the problem (\ref{MP2}) are solutions of the stationary, incompressible Euler equation, i.e., for any global minimiser $B$ there exists a smooth function $f$ with $B\times \text{curl}(B)=\text{grad}(f)$. In particular, global minimisers of (\ref{MP1}) turn out to be solutions of (\ref{MP2}), provided the minimiser is contained in the set $\mathcal{V}_{B_0}(\Omega)$. This is the content of \cref{C22}. \\
Beltrami fields are of particular interest from a topological point of view and have attracted a lot of interest in the mathematical community. It is shown in \cite{AK98} that nowhere vanishing steady flows on closed $3$-manifolds whose field lines are 'chaotic' are necessarily Beltrami-fields. On the other hand, in \cite{EP12} and \cite{EP15} the authors establish a result which states that for any prescribed (finite) collection of knots and links in $\mathbb{R}^3$ one can find a Beltrami field with field lines which - up to a diffeomorphism arbitrarily close to the identity in the $C^k$-norm- coincide with the given knots and links. Recently they generalised this result in \cite{EPT17} to the setting of the $3$-sphere and the $3$-torus. Sometimes eigenfields of the curl are called strong Beltrami-fields, while, more generally, any smooth vector field $X$ satisfying the equation $\text{curl}(X)=fX$ for a differentiable function $f$ is called a Beltrami-field. These type of vector fields have also been studied in recent years on $\mathbb{R}^3$ \cite{N14} and on open domains \cite{EP16}.
\section{Main results}
In this section we present our main results, show how they relate to already established results and explain the main ideas of the proofs of our theorems.
\newline
\newline
\textbf{General assumption:} For the rest of the paper we will assume that all manifolds $(\bar{M},g)$ in question are oriented, compact, smooth Riemannian $3$-manifolds with or without boundary and we will refer to manifolds with these properties simply as $3$-manifolds.
\newline
\newline
\textbf{Notation:} Let $(\bar{M},g)$ be a $3$-manifold. Then we denote by $\mathcal{V}(\bar{M})$ the set of all smooth vector fields on $\bar{M}$ and by $\mathcal{V}_n(\bar{M})$ the set of all smooth vector fields on $\bar{M}$, which admit a smooth vector potential normal to the boundary, that is, $X\in \mathcal{V}_n(\bar{M})$ if and only if there is some $A\in \mathcal{V}(\bar{M})$ with $\text{curl}(A)=X$ and $A\perp \partial\bar{M}$. Further we denote by $L^2\mathcal{V}_n(\bar{M})$ the completion of $\mathcal{V}_n(\bar{M})$ with respect to the $L^2$-norm induced by the metric $g$. We further define the helicity of a given $B\in L^2\mathcal{V}_n(\bar{M})$ by $\mathcal{H}(B):=\langle A,B\rangle_{L^2}$, where $A$ is any $H^1$-vector potential of $B$. This quantity is well-defined, i.e., independent of the choice of potential. Lastly consider the curl operator $\text{curl}:\mathcal{V}_n(\bar{M})\rightarrow \mathcal{V}(\bar{M})$, then we show that this operator admits a smallest positive and largest negative eigenvalue, which we denote by $\lambda_+>0$ and $\lambda_-<0$ respectively.
\newline
\newline
Let $(\bar{M},g)$ be a $3$-manifold. We consider the following minimisation problem for a fixed value $h\in \mathbb{R}$
\begin{gather}
\label{21}
\mathcal{E}:L^2\mathcal{V}_n(\bar{M})\rightarrow \mathbb{R},\text{ }B\mapsto \langle B,B\rangle_{L^2}, \quad \mathcal{E}(B)\rightarrowtail \text{min},\text{ }\mathcal{H}(B)=h.
\end{gather}
\begin{thm}[Main Theorem, Generalised Arnold's theorem]
	\label{T21}
	Let $(\bar{M},g)$ be a $3$-manifold, then the minimisation problem (\ref{21}) has a solution for every given $h\in \mathbb{R}$ and all minimisers are elements of $\mathcal{V}_n(\bar{M})$. In case of $h\neq 0$ also all local minimisers of $\mathcal{E}$ under the same constraint are elements of $\mathcal{V}_n(\bar{M})$ and they are Beltrami fields, that is, they are eigenvector fields of the curl operator corresponding to non-zero eigenvalues.
	\newline
	Moreover we have the following characterisation of global minimisers
	\newline
	\newline
	If $h= 0$, then $B=0$ is the unique global minimiser.
	\newline
	\newline
	If $h>0$, then $B\in L^2\mathcal{V}_n(\bar{M})$ is a global minimiser of the minimisation problem (\ref{21}) if and only if $B\in \mathcal{V}_n(\bar{M})$, $\mathcal{H}(B)=h$ and curl$(B)=\lambda_+B$.
	\newline
	\newline
	If $h<0$, then $B\in L^2\mathcal{V}_n(\bar{M})$ is a global minimiser of the minimisation problem (\ref{21}) if and only if $B\in \mathcal{V}_n(\bar{M})$, $\mathcal{H}(B)=h$ and curl$(B)=\lambda_-B$.
	\newline
	\newline
	If $g|_{\text{int}(\bar{M})}$ is real analytic, then local minimisers in non-trivial helicity classes are real analytic on $\text{int}(\bar{M})$.
\end{thm}
\textbf{Remarks:}
(i) By $g|_{\text{int}(\bar{M})}$ we denote the pullback of $g$ via the inclusion map and we call it real analytic if we can find a compatible real analytic atlas of $\text{int}(\bar{M})$ with respect to which $g|_{\text{int}(\bar{M})}$ is real analytic. The local minimisers are then real analytic with respect to the same analytic structure.
\newline
(ii) The space $\mathcal{V}_n(\bar{M})$ is infinite dimensional.
\newline
(iii) The set among which we wish to minimise the energy is always non-empty, that is, for all $h\in \mathbb{R}$: $\{X\in \mathcal{V}_n(\bar{M})|\mathcal{H}(X)=h\}\neq \emptyset$.
\newline
(iv) Minimisers of the minimisation problem (\ref{21}) are never unique, unless $h=0$, because we have the equalities $\mathcal{H}(-X)=\mathcal{H}(X)$ and $\mathcal{E}(-X)=\mathcal{E}(X)$. Thus whenever $X$ is a global minimiser, so is $-X$. Even modulo sign the solution is in general not unique because the eigenspaces corresponding to the eigenvalues $\lambda_+>0$ and $\lambda_-<0$ are in general not $1$-dimensional, see \cite{CDGT00}. However, the eigenspaces are always finite dimensional. In fact one only needs to slightly adjust the proof of \cite[Theorem 2.2.2]{S95} to obtain this result.
\newline
(v) Using the same notation as in \cref{T21} we have the following inequality for all $X\in L^2\mathcal{V}_n(\bar{M})$
\begin{gather}
\label{22}
\frac{1}{\lambda_-}\mathcal{E}(X)\leq \mathcal{H}(X)\leq \frac{1}{\lambda_+}\mathcal{E}(X)
\\
\label{23}
\text{and consequently if we set }\lambda:=\min\{|\lambda_-|,\lambda_+\}:\text{ }|\mathcal{H}(X)|\leq \frac{1}{\lambda}\mathcal{E}(X).
\end{gather}
Both inequalities are completely analogous to the inequalities Arnold obtains in \cite{A74} for manifolds without boundary.
\begin{cor}[Generalised Arnold's theorem for \ref{MP2}]
	\label{C22}
	Let $(\bar{M},g)$ be a $3$-manifold and let $B\in \mathcal{V}_n(\bar{M})$ satisfy either:
	\begin{gather}
	\label{24}
	\text{curl}(B)=\lambda_+B\quad\text{or} \quad \text{curl}(B)=\lambda_-B,
	\end{gather}
	then $B$ is an energy minimiser among the set of all the vector fields obtained from $B$ by the action of a volume-preserving diffeomorphism. That is, $B$ is a solution of the following minimisation problem
	\begin{gather}
	\label{25}
	\mathcal{E}:\mathcal{V}_B(\bar{M})\rightarrow \mathbb{R},\text{ }X\mapsto \langle X,X\rangle_{L^2}\rightarrowtail \min.
	\end{gather}
\end{cor}
Our final result is concerned with necessary conditions for general global minimisers of (\ref{MP2}) and is also a generalisation of a result by Arnold \cite{A74} for manifolds without boundary. To this end we define $\mathcal{V}_P(\bar{M})$ to be the set of all smooth, divergence-free vector fields which are tangent to the boundary of $\bar{M}$.
\begin{thm}[Euler-Lagrange equation for \ref{MP2}]
	\label{T23}
	Let $(\bar{M},g)$ be a $3$-manifold and let $B_0\in \mathcal{V}_P(\bar{M})$. Then we have the following inclusion
	\begin{gather}
	\label{26}
	\mathcal{V}_{B_0}(\bar{M})\subseteq\mathcal{V}_P(\bar{M}).
	\end{gather}
	That is, all vector fields obtained from $B_0$ by the action of a volume-preserving diffeomorphism are still divergence-free and tangent to the boundary. Furthermore any solution $B$ of the following minimisation problem
	\begin{gather}
	\label{27}
	\mathcal{E}:\mathcal{V}_{B_0}(\bar{M})\rightarrow \mathbb{R},\text{ }X\mapsto \langle X,X\rangle_{L^2}\rightarrowtail \min,
	\end{gather}
	is a solution of the stationary, incompressible Euler equation. That is, there exists a smooth function $f\in C^{\infty}(\bar{M})$, such that
	\begin{gather}
	\label{28}
	B\times \text{curl}(B)=\text{grad}(f).
	\end{gather}
\end{thm}
In his paper \cite{A74} Arnold essentially proves the analogous results for closed manifolds. Setting $\partial \bar{M}=\emptyset$ in \cref{T23} exactly reproduces Arnold's result as a special case. The main idea of the proof of \cref{T23} can be carried over from Arnold's proof. The only obstacle we face is that we need to deal with boundary terms, which, however, will turn out to vanish. If we let $\partial\bar{M}=\emptyset$ in \cref{C22}, then we obtain a seemingly stronger result than Arnold, because Arnold states his result only for the special case $\text{curl}(B)=\lambda B$, where $\lambda \in \{\lambda_-,\lambda_+ \}$ is the eigenvalue of smallest modulus. However, Arnold's reasoning can be used to derive the same result obtained from \cref{C22} for $\partial \bar{M}=\emptyset$, i.e., Arnold unnecessarily restricts himself to the eigenvalue of smallest modulus. The idea of our proof is the same as Arnold's. We show that the helicity is conserved under the action of volume-preserving diffeomorphisms and then apply \cref{T21} to derive \cref{C22}.
\newline
Let us now comment on the proof of \cref{T21}. Here our approach differs from Arnold's spectral theoretical approach. We instead use a Lagrange multiplier method inspired by Woltjer's original work \cite{W58}, where he formally shows that in IMHD certain Beltrami fields are local magnetic energy minimisers in closed physical systems. We will make this idea rigorous for our setting and in return obtain a result which does not only reveal the Beltrami field property of global, but instead of all local minimisers. Hence even for the case $\partial \bar{M}=\emptyset$ our result is a strict generalisation of Arnold's result. The characterisation of global minimisers, as given in \cref{T21}, can neither be found explicitly in \cite{A74} nor in \cite{AK98}. But the spectral theoretical approach allows for such a characterisation, so that the corresponding result is already implicitly contained in Arnold's work.
\newline
In fact the Lagrange multiplier approach was already made rigorous in \cite{AL91}. Therein the authors consider a more general class of boundary conditions and work with the notion of relative helicity, which reduces to our definition of helicity for our choice of boundary conditions. However, the authors in \cite{AL91} entirely restrict themselves to the case of Euclidean domains and in order to justify the usage of a multiplier approach they derive a corresponding version for the particular problem at hand. On the contrary we work in the more general abstract manifold setting and only assume orientability and compactness of the underlying space. In addition we utilise an abstract Lagrange multiplier method for Banach spaces in order to make Woltjer's idea rigorous. In this sense our approach differs from theirs.
\newline
As for the regularity one observes that each Beltrami field $X$ satisfies the equation $-\Delta X=\lambda^2 X$ and hence in the Euclidean setting each component is an eigenfunction of the Laplacian. Hence real analyticity of smooth eigenfunctions follows. In our setting we get a coupled system of equations which require a more refined analysis. Nonetheless elliptic estimates are still available which allow us to conclude the desired regularity result. Let us lastly point out that the authors in \cite{AL91} solely deal with the minimisation problem (\ref{MP1}) so that in this sense our paper gives a more complete picture of the problem as a whole.
\newline
There are also some other established results concerning manifolds with boundary, see for instance \cite{AK98} and \cite{CDG00}. In \cite{AK98} Arnold and Khesin consider the case of simply connected bounded domains $\Omega\subset \mathbb{R}^3$ with smooth boundary. Therein they define the helicity for any divergence-free smooth vector field, which is tangent to the boundary of $\Omega$. They show that in this case the helicity is also independent of the choice of vector potential and use a spectral theoretical argument as in \cite{A74} to derive a result corresponding to our \cref{C22} for this special case (even though they again restrict themselves to the case of the eigenvalue of smallest modulus). They comment in a short remark, without providing a proof, that this result generalises to compact, smooth, Riemannian $3$-manifolds with boundary with vanishing first de Rham cohomology. We recall that in our case we define the helicity on the set $\mathcal{V}_n(\bar{M})$. In fact we always have the inclusion $\mathcal{V}_n(\bar{M})\subseteq \mathcal{V}_P(\bar{M})$, meaning that every smooth vector field admitting a vector potential, which is normal to the boundary, is divergence-free and tangent to the boundary. However equality holds if and only if the first de Rham cohomology of $\bar{M}$ vanishes
\begin{prop}
	\label{P24}
	Let $(\bar{M},g)$ be a $3$-manifold. Then the following holds
	\begin{gather}
	\label{29}
	\mathcal{V}_n(\bar{M})=\mathcal{V}_P(\bar{M}) \Leftrightarrow H^1_{dR}(\bar{M})=\{0\}.
	\end{gather}
\end{prop}
In other words, if the first de Rham cohomology of the underlying manifold vanishes, the helicity in our case is also defined on the set of all divergence-free vector fields tangent to the boundary and coincides with the notion of helicity introduced in \cite{AK98}. Hence as a special case we provide a proof of the remark from \cite{AK98}. In \cite{CDG00} Cantarella, DeTurck and Gluck give a definition of helicity on arbitrary bounded domains $\Omega\subset \mathbb{R}^3$ with smooth boundary for smooth vector fields which are divergence-free and tangent to the boundary of $\Omega$. The helicity they define can be expressed as the $L^2$-inner product of the vector field in question and its Biot-Savart potential. For general, non-simply connected domains, the value of the $L^2$-inner products of the vector field and its vector potentials will depend on the choice of potential. Therefore if we wish to express the helicity as the $L^2$-inner product of the vector field and one of its vector potentials, it is essential to assign a specific vector potential to every divergence-free vector field, which is tangent to the boundary. Using a spectral theoretical approach Cantarella et al. show that the strongly related problem of maximising the helicity for prescribed energy admits a solution and that such solutions are Beltrami fields. Note that we in general do not reproduce the results from \cite{CDG00}, unless $H^1_{dR}(\Omega)=\{0\}$. Because only in this case our notion of helicity is defined for all divergence-free vector fields tangent to the boundary. This is due to the fact that we do not pick any specific potential for the definition of helicity, but instead restrict ourselves to a set of vector fields for which the helicity is independent of a particular choice of potential. In case of $H^1_{dR}(\Omega)=\{0\}$ our results coincide, but in general, if $H^1_{dR}(\Omega)\neq \{0\}$, there is a priori no reason to believe that our solutions will also be solutions of the problem studied by Cantarella et al., because the set over which they maximise helicity in these cases is larger. However they show that the solutions to their problem are Beltrami fields corresponding to the largest positive eigenvalue of a modified Biot-Savart operator. If we let $\mu_+>0$ denote this largest eigenvalue and $\lambda_+>0$ denote the smallest positive eigenvalue of the curl operator as in \cref{T21}, then we always have the inequality $\frac{1}{\mu_+}\leq \lambda_+$ with equality if and only if there is a solution of their problem contained in $\mathcal{V}_n(\bar{\Omega})$.
\newline
To the best of my knowledge, even though the results presented in this paper are known among experts, there is no citable source concerning the existence of minimisers on arbitrary abstract, compact manifolds with boundary as described in \cref{T21}. The aim of this paper is to fill this gap.
\section{Preliminary results}
In view of the fact that the Hodge-Morrey decomposition theorem and integration theory is formulated for forms, rather than for vector fields, we will be working on the level of $1$-forms instead of working with vector fields directly. There is a canonical way to identify smooth vector fields and (smooth) $1$-forms via the Riemannian metric of the manifold. We will denote the isomorphism between these spaces by $\omega^1$ and for any smooth vector field $X$ we denote by $\omega^1_X$ the corresponding $1$-form. Observe that this isomorphism in fact induces an $L^2$-isometry between these spaces and hence extends uniquely to an isometry between the $L^2$-completions of these spaces. In addition we can write the energy as $\mathcal{E}(X)=\langle X,X\rangle_{L^2}=(\omega^1_X,\omega^1_X)_{L^2}$ (the different brackets indicate the different $L^2$-inner products on the respective spaces), and the helicity is also defined in terms of the $L^2$-inner product. Thus we see that there will be an immediate correspondence between the results we derive for $1$-forms and the results for vector fields. For an introduction to basic concepts, Sobolev theory and the Hodge-Morrey-Friedrichs decomposition on abstract manifolds with boundary we refer the reader to \cite{S95}. We denote by $t(\omega)$, $n(\omega)$ the tangent and normal part of a $k$-form $\omega$, respectively, and by $\mathcal{V}(\bar{M})$, $\Omega^k(\bar{M})$ the spaces of all smooth vector fields and (smooth) $k$-forms on $\bar{M}$ respectively. We also repeatedly use the Hodge-Morrey-Friedrichs decomposition \cite[Theorem 2.4.2,Theorem 2.4.8]{S95} and Green's formula/integration by parts formula \cite[Proposition 2.1.2]{S95}, which the reader is assumed to be familiar with. Lastly if $\mathcal{S}\subseteq \Omega^k(\bar{M})$ and $s\in \mathbb{N}_0$ we will always denote by $H^s\mathcal{S}$ the completion of $\mathcal{S}$ with respect to the Sobolev-norm $\norm{\cdot}_{H^s}$ \cite[Chapter 1.3]{S95}. As usual we identify $H^0=L^2$.
\newline
Let us shortly recall here that for every $p\in \partial\bar{M}$ we can decompose every tangent vector $V\in T_p\bar{M}$ uniquely into a part $V^{\perp}$, normal to the boundary, and a part $V^{\parallel}$, tangent to the boundary, such that $V=V^{\parallel}+V^{\perp}$. The tangent part of a $k$-form is then defined as $t(\omega)(p)(V_1,\dots,V_k):=\omega(p)(V^{\parallel}_1,\dots,V^{\parallel}_k)$ and $n(\omega)(p):=\omega(p)-t(\omega)(p)$ for $p\in \partial\bar{M}$ and $V_i\in T_p\bar{M}$. In particular we have $X^{\parallel}=0\Leftrightarrow t(\omega^1_X)=0$ and $X^{\perp}=0\Leftrightarrow n(\omega^1_X)=0$.
\begin{lem}
	\label{L31}
	Let $(\bar{M},g)$ be a $3$-manifold and let $\mathcal{V}_n(\bar{M})$ denote the set of all smooth vector fields $X$ on $\bar{M}$, which admit a smooth vector potential $A$ normal to the boundary. Then
	\begin{gather}
	\label{31}
	\omega^1\left(\mathcal{V}_n(\bar{M}) \right)=\{\omega\in \Omega^1(\bar{M})| \omega=\delta \Omega\text{ for some }\Omega\in \Omega^2(\bar{M})\text{ with }n(\Omega)=0 \},
	\end{gather}
	where $\delta$ denotes the adjoint derivative. We denote this space by $\Omega^1_n(\bar{M})$.
\end{lem}
\underline{Proof:} The proof is straightforward. One only needs to keep in mind that $A^{\parallel}=0$ is equivalent to $t(\omega^1_A)=0$ and in addition we have the duality relations $\star t(\omega)=n(\star \omega)$ and $\star n(\omega)=t(\star \omega)$ for all forms $\omega$, \cite[Proposition 1.2.6]{S95}. $\square$
\begin{defn}[Helicity]
	\label{D32}
	Let $(\bar{M},g)$ be a $3$-manifold. We define the \textbf{helicity} on $\Omega^1_n(\bar{M})$ via
	\begin{gather}
	\label{32}
	\mathcal{H}:\Omega^1_n(\bar{M})\rightarrow \mathbb{R}, \omega\mapsto \left(\omega,\star \tilde{\Omega} \right)_{L^2},
	\end{gather}
	where $\tilde{\Omega}\in \Omega^2(\bar{M})$ is any smooth $2$-form satisfying $\omega=\delta \tilde{\Omega}$.
\end{defn}
\begin{lem}
	\label{L33}
	The helicity in the setting of \cref{D32} is well-defined, that is the inner product is independent of a particular choice of potential $\tilde{\Omega}$.
\end{lem}
\underline{Proof:} Let $\tilde{\Omega}\in \Omega^2(\bar{M})$ be any $2$-form with $\omega=\delta \tilde{\Omega}$. By definition of the space $\Omega^1_n(\bar{M})$ we can find another $2$-form $\Omega$ with $\omega=\delta \Omega$ and $n(\Omega)=0$. We compute via Green's formula \cite[Proposition 2.1.2]{S95} and the $L^2$-isometry of $\star$
\[
\left(\omega,\star \tilde{\Omega}-\star \Omega \right)_{L^2}=\left(\delta \Omega,\star \tilde{\Omega}-\star \Omega \right)_{L^2}=\left(\star \Omega,\delta \tilde{\Omega}-\delta \Omega \right)_{L^2}=0,
\]
where the boundary term, which usually appears, vanishes because $n(\Omega)=0$ and the last equality holds because $\delta\Omega=\omega=\delta \tilde{\Omega}$. $\square$
\begin{lem}
	\label{L34}
	Let $(\bar{M},g)$ be a $3$-manifold. Define the space
	\begin{gather}
	\label{33}
	\Omega^1_T(\bar{M}):=\{\alpha\in \Omega^1(\bar{M})|t(\alpha)=0\text{ and }\exists \Omega\in \Omega^2(\bar{M})\text{ with }\delta\Omega=\alpha \}.
	\end{gather}
	Then the following operator is bijective
	\begin{gather}
	\label{34}
	\text{curl}:\Omega^1_T(\bar{M})\rightarrow \Omega^1_n(\bar{M}),\alpha\mapsto \star d\alpha.
	\end{gather}
	We will denote its inverse by $\text{curl}^{-1}:\Omega^1_n(\bar{M})\rightarrow \Omega^1_T(\bar{M})$, or shortly by curl$^{-1}$.
\end{lem}
\underline{Proof of \cref{L34}:} The map is obviously well-defined because for $\alpha\in \Omega^1_T(\bar{M})$ we have $\star d \alpha=\delta(\star \alpha)$ and $n(\star \alpha)=\star t(\alpha)=0$ by the duality relation and by definition of $\Omega^1_T(\bar{M})$.
\newline
\newline
\underline{injective:} By linearity it is enough to show that the kernel is trivial. So let $\delta\Omega\in \Omega^1_T(\bar{M})$ satisfy curl$(\delta\Omega)=0\Leftrightarrow d\delta \Omega=0$. Then we have by Green's formula $(\delta \Omega,\delta\Omega)_{L^2}=(\Omega,d\delta \Omega)_{L^2}=0$, where we used that $t(\delta\Omega)=0$ by definition of $\Omega^1_T(\bar{M})$ and hence the boundary term vanishes. Thus $\delta\Omega=0$ as desired.
\newline
\newline
\underline{surjective:} Let $\omega\in \Omega^1_n(\bar{M})$, then by definition there exists an $\Omega\in \Omega^2(\bar{M})$ with $n(\Omega)=0$ and $\omega=\delta \Omega$. Note that by the duality relation we have $t(\star \Omega)=0$ and $\omega=\text{curl}(\star \Omega)$. On the other hand applying the Hodge-Morrey-Friedrichs decomposition \cite[Theorem 2.4.2,Theorem 2.4.8]{S95}, we find smooth forms $\alpha,\beta,\gamma$ of appropriate degree such that $\star \Omega=d\alpha+\delta \beta+\gamma$, $t(d\alpha)=0$, $t(\gamma)=0$ and $d\gamma=0=\delta\gamma$. By linearity of the operator $t$ we find $0=t(\star\Omega)=t(d\alpha)+t(\delta\beta)+t(\gamma)=t(\delta\beta)$. By definition we find $\delta \beta\in \Omega^1_T(\bar{M})$ and we compute $\text{curl}(\delta \beta)=\star d(\delta \beta)=\star d (d\alpha+\delta\beta+\gamma)=\text{curl}( \star \Omega)=\omega$, where we used that $d\gamma=0=d^2\alpha$. $\square$
\begin{lem}
	\label{L35}
	Let $(\bar{M},g)$ be a $3$-manifold. Then the inverse curl operator, as defined in \cref{L34}, extends to a continuous linear operator
	\begin{gather}
	\label{35}
	\text{curl}^{-1}:\left(L^2\Omega^1_n(\bar{M}),\norm{\cdot}_{L^2} \right)\rightarrow \left(H^1\Omega^1_T(\bar{M}),\norm{\cdot}_{H^1} \right),
	\end{gather}
	and to a linear compact operator
	\begin{gather}
	\label{36}
	\text{curl}^{-1}:\left(L^2\Omega^1_n(\bar{M}),\norm{\cdot}_{L^2} \right)\rightarrow \left(H^1\Omega^1_T(\bar{M}),\norm{\cdot}_{L^2} \right),
	\end{gather}
	where $L^2\Omega^1_n(\bar{M})$ denotes the $L^2$-completion of $\Omega^1_n(\bar{M})$ and $H^1\Omega^1_T(\bar{M})$ denotes the $H^1$-completion of $\Omega^1_T(\bar{M})$.
\end{lem}
\underline{Proof:} Once we have established (\ref{35}) it will imply (\ref{36}) by standard arguments in combination with the fact that the inclusion $\iota: \left(H^1\Omega^1(\bar{M}),\norm{\cdot}_{H^1}\right)\rightarrow \left(L^2\Omega^1(\bar{M}),\norm{\cdot}_{L^2} \right)$ is compact by the Sobolev embedding theorem \cite[Theorem 1.3.6]{S95}. In order to see (\ref{35}) we observe that for $\omega\in \Omega^1_n(\bar{M})$ we have curl$^{-1}(\omega)\in \Omega^1_T(\bar{M})$ and so in particular $t(\text{curl}^{-1}(\omega))=0$. On the other hand curl$^{-1}(\omega)$ is $L^2$-orthogonal to the space of harmonic Dirichlet fields $\mathcal{H}^1_D(\bar{M}):=\{\gamma\in \Omega^1(\bar{M})|d\gamma=0=\delta\gamma\text{ and }t(\gamma)=0 \}$. To see this let $\gamma\in \mathcal{H}^1_D(\bar{M})$ and recall that by definition of $\Omega^1_T(\bar{M})$ we can write curl$^{-1}(\omega)=\delta\Omega$ for a suitable $\Omega\in \Omega^2(\bar{M})$. Hence we obtain by Green's formula
\[
(\text{curl}^{-1}(\omega),\gamma)_{L^2}=(\delta \Omega,\gamma)_{L^2}=(\Omega,d\gamma)_{L^2}=0,
\]
where the boundary term vanishes because $t(\gamma)=0$. Then \cite[Lemma 2.4.10]{S95} implies that there exists a constant $C>0$ (independent of $\omega$) such that the following estimate holds
\[
\norm{\text{curl}^{-1}(\omega)}_{H^1}\leq C \left(\norm{d (\text{curl}^{-1}(\omega))}_{L^2}+\norm{\delta (\text{curl}^{-1}(\omega))}_{L^2}\right).
\]
We observe that $\delta (\text{curl}^{-1}(\omega))=\delta^2\Omega=0$ , that $\star$ is an $L^2$-isometry and that $\star d=\text{curl}$. This yields
\[
\norm{\text{curl}^{-1}(\omega)}_{H^1}\leq C\norm{\omega}_{L^2}.
\]
This proves continuity of the operator on the underlying spaces $\Omega^1_n(\bar{M})$ and $\Omega^1_T(\bar{M})$. Standard arguments from functional analysis imply that there exists a unique continuous extension to the respective completions of the corresponding spaces. $\square$
\begin{flushleft}
	\textbf{Remark:} (i) Notice that the curl operator $\text{curl}:\left(\Omega^1_T(\bar{M}),\norm{\cdot}_{H^1} \right)\rightarrow \left(\Omega^1_n(\bar{M}),\norm{\cdot}_{L^2} \right)$ is also bounded and hence extends to a continuous operator $\text{curl}:\left(H^1\Omega^1_T(\bar{M}),\norm{\cdot}_{H^1} \right)\rightarrow \left(L^2\Omega^1_n(\bar{M}),\norm{\cdot}_{L^2} \right)$ on the respective completions of the underlying spaces. A standard density argument implies that this extended curl operator and the operator curl$^{-1}$ from \cref{L35} are inverses of one another.
	\newline
	(ii) In \cite[Lemma 1]{YG90} the authors introduce similar boundary conditions in the Euclidean setting and prove the existence of a compact inverse curl operator in that setting.
\end{flushleft}
\begin{defn}[Helicity on $L^2\Omega^1_n(\bar{M})$]
	\label{D36}
	Let $(\bar{M},g)$ be a $3$-manifold. We define the \textbf{helicity} on $L^2\Omega^1_n(\bar{M})$ via
	\begin{gather}
	\label{37}
	\mathcal{H}:L^2\Omega^1_n(\bar{M})\rightarrow \mathbb{R},\omega\mapsto \left(\omega,\text{curl}^{-1}(\omega)\right)_{L^2},
	\end{gather}
	where curl$^{-1}$ is the extended inverse operator from \cref{L35}. For $\omega\in \Omega^1_n(\bar{M})$ this coincides with \cref{D32} because curl$^{-1}(\omega)$ is a vector potential of $\omega$.
\end{defn}
\textbf{Remark:} We use here an explicit vector potential $\text{curl}^{-1}(\omega)\in H^1\Omega^1_T(\bar{M})$ of $\omega\in L^2\Omega^1_n(\bar{M})$ to define its helicity. But one can show that just like in the smooth case, if $\alpha\in H^1\Omega^1(\bar{M})$ is any other $H^1$-$1$-form with $\star d\alpha=\omega$, then $\left(\alpha,\omega\right)_{L^2}=\mathcal{H}(\omega)$. Thus the helicity is still independent of a particular choice of potential.
\section{Proofs of main results}
\subsection{Proof of theorem 2.1}
\underline{Step 1: Existence of global minimisers}
\newline
\newline
In the first step we will use the direct method in calculus of variations to show that the minimisation problem (\ref{21}) admits a global minimiser for all $h\in \mathbb{R}$. To this end let $(\omega_k)_k\subset L^2\Omega^1_n(\bar{M})$ be a minimising sequence. We observe that in particular $(\omega_k)_k$ is $L^2$-bounded and since $L^2\Omega^1_n(\bar{M})$ is a Hilbert space there exists some $\omega \in L^2\Omega^1_n(\bar{M})$ such that $\omega_k \rightharpoonup \omega$ weakly in $L^2$ for $k\rightarrow \infty$ (after extracting a subsequence if necessary). Obviously the square of the $L^2$-norm is $L^2$-weakly lower semi-continuous, so that $\omega$ will turn out to be a global minimiser once we show that $\mathcal{H}(\omega)=h$. However we also know that due to the compactness of the operator in (\ref{36}) we may assume (after extracting yet another subsequence) that $(\text{curl}^{-1}(\omega_k))_k\subset H^1\Omega^1_T(\bar{M})$ converges strongly in $L^2$. It is now a standard task to confirm that $\mathcal{H}(\omega)=\lim_{k\rightarrow \infty}\mathcal{H}(\omega_k)=h$. $\square$
\newline
\newline
\underline{Step 2: Regularity of local minimisers and the Beltrami field property}
\newline
\newline
To keep the regularity proof more accessible we divide it into two parts which we will state as separate lemmas.
\begin{lem}
	\label{L41}
	Suppose we are in the setting of \cref{T21} and let $\omega\in L^2\Omega^1_n(\bar{M})$ be a local minimiser of (\ref{21}) for a $h\in \mathbb{R}\setminus \{0\}$. Then $\omega \in L^2\Omega^1_n(\bar{M})\cap H^1\Omega^1(\bar{M})$ and there exists a constant $\lambda_{\omega}\in \mathbb{R}\setminus \{0\}$ such that
	\begin{gather}
	\label{41}
	\text{curl}(\omega)=\lambda_{\omega}\omega.
	\end{gather}
\end{lem}
\underline{Proof of \cref{L41}:} In order to derive this result we wish to apply the Lagrangian multiplier method for Banach spaces. To this end we need to check that the energy functional, as well as the constraint function $\mathcal{H}$ are both continuously $L^2$-Fr\'{e}chet differentiable. It is straightforward to check that this is the case, keeping in mind the continuity of curl$^{-1}$. We compute that the derivatives of $\mathcal{E}$ and $\mathcal{H}$ at a point $\alpha\in L^2\Omega^1_n(\bar{M})$ are given by
\[
\mathcal{E}^{\prime}(\alpha)(\phi)=2\left(\alpha,\phi\right)_{L^2}\text{ and }\mathcal{H}^{\prime}(\alpha)(\phi)=\left(\alpha,\text{curl}^{-1}(\phi)\right)_{L^2}+\left(\phi,\text{curl}^{-1}(\alpha)\right)_{L^2}
\]
for $\phi\in L^2\Omega^1_n(\bar{M})$. Using an approximation argument and Green's formula one concludes that $\left(\alpha,\text{curl}^{-1}(\phi)\right)_{L^2}$ $=\left(\phi,\text{curl}^{-1}(\alpha)\right)_{L^2}$ and thus $\mathcal{H}^{\prime}(\alpha)(\phi)=2\left(\phi,\text{curl}^{-1}(\alpha)\right)_{L^2}$. In order to apply the Lagrangian multiplier method we need to check that $\mathcal{H}^{\prime}(\omega)$ is surjective. Since $\mathcal{H}^{\prime}(\omega)$ maps into the real numbers it is enough to show that there is some $\phi\in L^2\Omega^1_n(\bar{M})$ with $\mathcal{H}^{\prime}(\omega)(\phi)\neq 0$. But by assumption $\omega\in L^2\Omega^1_n(\bar{M})$ is a local minimiser of (\ref{21}) for some fixed $h\neq 0$. Thus we may choose $\phi=\omega$ and find that $\mathcal{H}^{\prime}(\omega)(\omega)=2\mathcal{H}(\omega)=2h\neq 0$. Hence we may apply the Lagrangian multiplier method \cite[p.270 Proposition 1]{Z95} to conclude that there is some $\lambda\in \mathbb{R}$ with
\begin{gather}
\label{42}
\left(\omega-\lambda \text{ } \text{curl}^{-1}(\omega),\phi \right)_{L^2}=0\text{ for all }\phi \in L^2\Omega^1_n(\bar{M}).
\end{gather}
Note that $\lambda \neq 0$ because otherwise we may insert $\phi=\omega$ in (\ref{42}) to conclude that $\omega=0$, which contradicts $\mathcal{H}(\omega)=h\neq 0$. By definition we may now approximate $\omega$ by a sequence $(\omega_k)_k\subset \Omega^1_n(\bar{M})$ in $L^2$-norm. Consequently by \cref{L35} $(\text{curl}^{-1}(\omega_k))_k\subset \Omega^1_T(\bar{M})$ converges to curl$^{-1}(\omega)$ in $H^1$- and hence in particular in $L^2$-norm. We recall that by definition of the space $\Omega^1_T(\bar{M})$ we can find $\Omega_k\in \Omega^2(\bar{M})$ such that $\text{curl}^{-1}(\omega_k)=\delta \Omega_k$. In view of this and the Hodge-Morrey-Friedrichs decomposition we can decompose curl$^{-1}(\omega_k)$ as
\begin{gather}
\label{43}
\text{curl}^{-1}(\omega_k)=\delta \beta_k+\delta \gamma_k, \text{ for }\beta_k,\gamma_k\in \Omega^2(\bar{M})\text{ with }n(\beta_k)=0\text{ and }d\delta\gamma_k=0.
\end{gather}
By the $L^2$-orthogonality of this decomposition we have the following equality for all $k,m\in \mathbb{N}$
\[
\norm{\delta\beta_k-\delta\beta_m}^2_{L^2}+\norm{\delta\gamma_k-\delta\gamma_m}^2_{L^2}=\norm{\text{curl}^{-1}(\omega_k)-\text{curl}^{-1}(\omega_m)}^2_{L^2}.
\]
But $(\text{curl}^{-1}(\omega_k))_k$ converges strongly in $L^2$ to $\text{curl}^{-1}(\omega)$ so that it is an $L^2$-Cauchy sequence. We conclude that $(\delta\beta_k)_k$ and $(\delta\gamma_k)_k$ are both $L^2$-Cauchy sequences and converge to some $\eta,\tilde{\gamma}\in L^2\Omega^1(\bar{M})$ respectively. We observe that $n(\beta_k)=0$ implies that $\delta \beta_k\in \Omega^1_n(\bar{M})$ for all $k$ and hence we conclude by definition that $\eta=\lim_{k\rightarrow\infty}\delta\beta_k\in L^2\Omega^1_n(\bar{M})$. An approximation argument and Green's formula yield $(\tilde{\gamma},\phi)_{L^2}=0$ for all $\phi\in L^2\Omega^1_n(\bar{M})$ and in addition we conclude from (\ref{43}) that $\text{curl}^{-1}(\omega)=\eta+\tilde{\gamma}$. Plugging in our considerations so far in (\ref{42}) we obtain
\[
\left(\omega-\lambda\text{ } \eta,\phi \right)_{L^2}=0\text{ for all }\phi\in L^2\Omega^1_n(\bar{M}).
\]
Recall that $\omega$ and $\eta$ are elements of $L^2\Omega^1_n(\bar{M})$ so that we may set $\phi=\omega-\lambda\eta$ to conclude
\begin{gather}
\label{44}
\omega=\lambda\eta.
\end{gather}
We will now show that the sequence $(\delta\beta_k)_k$ is an $H^1$-Cauchy sequence, which will imply that $\eta \in L^2\Omega^1_n(\bar{M})\cap H^1\Omega^1(\bar{M})$ because $(\delta\beta_k)_k$ already converges to $\eta$ in $L^2$. This will prove the regularity assertions of \cref{L41}. Similarly to our proof of \cref{L35} we will use a suitable elliptic estimate to establish the Cauchy sequence property. The following elliptic estimate holds \cite[Lemma 2.4.10]{S95}: There is a constant $C>0$ such that for all $\alpha\in \Omega^1(\bar{M})$ with $\left(\alpha,\hat{\gamma}\right)_{L^2}=0$ for all $\hat{\gamma}\in \mathcal{H}^1(\bar{M}):=\{\gamma\in H^1\Omega^1(\bar{M})|d\gamma=0=\delta\gamma \}$ we have the estimate
\[
\norm{\alpha}_{H^1}\leq C\left(\norm{d\alpha}_{L^2}+\norm{\delta \alpha}_{L^2}\right).
\]
However the $L^2$-orthogonality of $\delta\beta_k-\delta\beta_m$ for any fixed indices $k,m\in \mathbb{N}$ to $\mathcal{H}^1(\bar{M})$ is a direct consequence of Green's formula and the fact that $n(\beta_k)=0=n(\beta_m)$. We can now argue, keeping in mind (\ref{43}), that $\delta(\delta\beta_k-\delta\beta_m)=0$ and that $\star d\delta\beta_k=\star d(\delta\beta_k+\delta\gamma_k)=\star d (\text{curl}^{-1}(\omega_k))=\omega_k$ because $d\delta\gamma_k=0$ and $\star d=\text{curl}$. Using the fact that $\star$ is an $L^2$-isometry we obtain the estimate
\begin{gather}
\label{45}
\norm{\delta\beta_k-\delta\beta_m}_{H^1}\leq C \norm{\omega_k-\omega_m}_{L^2}.
\end{gather}
However by choice of our sequence $(\omega_k)_k$ we know that it converges strongly in $L^2$ to $\omega$. Therefore it is an $L^2$-Cauchy sequence and so by (\ref{45}) $(\delta\beta_k)_k$ defines an $H^1$-Cauchy sequence, which implies $\eta \in H^1\Omega^1(\bar{M})$. This in combination with (\ref{44}) yields $\omega\in L^2\Omega^1_n(\bar{M})\cap H^1\Omega^1(\bar{M})$. Since we established enough regularity we may apply the curl operator on both sides of (\ref{44}) and arrive at
\begin{gather}
\label{46}
\text{curl}(\omega)=\lambda\text{ curl}(\eta).
\end{gather}
We lastly claim that curl$(\eta)=\omega$, which in combination with our observation that $\lambda\neq 0$ will conclude the proof of the lemma. As we have seen $(\delta \beta_k)_k$ is an $H^1$-Cauchy sequence and converges strongly in $L^2$ to $\eta$ and hence it converges strongly in $H^1$ to $\eta$. This implies that $(\text{curl}(\delta \beta_k))_k$ converges strongly in $L^2$ to $\text{curl}(\eta)$. But as we have argued before we have $\text{curl}(\delta \beta_k)=\star d \delta\beta_k=\omega_k$ and by choice of our sequence $(\omega_k)_k$ it converges strongly in $L^2$ to $\omega$. Therefore $\text{curl}(\eta)=\lim_{k\rightarrow \infty}\text{curl}(\delta \beta_k)=\lim_{k\rightarrow\infty}\omega_k=\omega$. $\square$
\begin{lem}
	\label{L42}
	Let $(\bar{M},g)$ be a $3$-manifold. Suppose $\omega\in L^2\Omega^1_n(\bar{M})\cap H^1\Omega^1(\bar{M})$ satisfies
	\begin{gather}
	\label{47}
	\text{curl}(\omega)=\lambda_{\omega}\omega,
	\end{gather}
	for some constant $\lambda_{\omega}\neq 0$. Then $\omega\in \Omega^1_n(\bar{M})$.
\end{lem}
\underline{Proof of \cref{L42}:} 
\newline
We recall that $\mathcal{H}^1_D(\bar{M})=\{\gamma\in \Omega^1(\bar{M})|d\gamma=0=\delta\gamma\text{ and }t(\gamma)=0 \}$ and we let $\mathcal{H}^1_D(\bar{M})^{\perp}$ denote its $L^2$-orthogonal complement. Furthermore we define the space $H^1\Omega^1_D(\bar{M}):=\{\alpha\in H^1\Omega^1(\bar{M})|t(\alpha)=0 \}$. Then according to \cite[Theorem 2.2.4]{S95} there exists for every $\xi\in \mathcal{H}^1_D(\bar{M})^{\perp}$ a unique element $\phi_D\in \mathcal{H}^1_D(\bar{M})^{\perp}\cap H^1\Omega^1_D(\bar{M})$, the so called Dirichlet potential of $\xi$, which is uniquely determined by the equation
\begin{gather}
\label{48}
\left(d\phi_D,d\eta\right)_{L^2}+\left(\delta \phi_D,\delta \eta\right)_{L^2}=\left(\xi,\eta\right)_{L^2}\text{ for all }\eta\in H^1\Omega^1_D(\bar{M}).
\end{gather}
We will now show in a first step that $\text{curl}^{-1}(\omega)\slash \lambda_{\omega}$ is the Dirichlet potential of $\omega$. First of all we recall that by definition of the space $L^2\Omega^1_n(\bar{M})$ we can approximate $\omega$ in $L^2$ by a sequence $(\delta\Omega_k)_k\subset \Omega^1_n(\bar{M})$ with $n(\Omega_k)=0$ for all $k$. It is then a direct consequence of Green's formula that $\omega\in \mathcal{H}^1_D(\bar{M})^{\perp}$. This implies that $\omega$ admits a Dirichlet potential. On the other hand we recall that $\text{curl}^{-1}(\omega)\in H^1\Omega^1_T(\bar{M})$. An approximation argument in combination with the trace theorem \cite[Theorem 1.3.7]{S95} implies that $t(\text{curl}^{-1}(\omega))=0$ and hence overall $\text{curl}^{-1}(\omega)\in H^1\Omega^1_D(\bar{M})$. Similarly we conclude $\text{curl}^{-1}(\omega)\in \mathcal{H}^1_D(\bar{M})^{\perp}$. Therefore if we can show that $\text{curl}^{-1}(\omega)\slash \lambda_{\omega}$ satisfies (\ref{48}), \cite[Theorem 2.2.4]{S95} will imply that it coincides with the Dirichlet potential $\phi_D$ of $\omega$. To see that (\ref{48}) is satisfied we first observe that by definition of the space $H^1\Omega^1_T(\bar{M})$ we have $\delta (\text{curl}^{-1}(\omega))=0$. On the other hand since $\star$ defines an $L^2$-isometry and since curl$(\text{curl}^{-1}(\omega))=\omega$, we have
\[
\left(d(\text{curl}^{-1}(\omega)),d\eta\right)_{L^2}=\left(\omega,\star d\eta\right)_{L^2}=\left(\star \omega,d\eta\right)_{L^2}=\left(\text{curl}(\omega),\eta\right)_{L^2}=\lambda_{\omega}\left(\omega,\eta\right)_{L^2},
\]
where we used Green's formula, the boundary condition $t(\eta)=0$ since $\eta \in H^1\Omega^1_D(\bar{M})$ and the Beltrami field property of $\omega$ in the last step. Since $\lambda_{\omega}\neq 0$ we may divide both sides by $\lambda_{\omega}$ and combining all our considerations so far we conclude that $\text{curl}^{-1}(\omega)\slash \lambda_{\omega}$ is the Dirichlet potential of $\omega$.
\newline
The Dirichlet potential has a well-established regularity theory \cite[Theorem 2.2.6]{S95}. In particular if $\omega\in H^k\Omega^1(\bar{M})$, then the corresponding Dirichlet potential $\phi_D$ satisfies $\phi_D\in H^{k+2}\Omega^1(\bar{M})$. Observe that $\text{curl}^{-1}(\omega)$ differs by the Dirichlet potential of $\omega$ only by a constant factor, so that the regularity result immediately carries over to $\text{curl}^{-1}(\omega)$. A standard bootstrapping argument implies that $\omega \in H^k\Omega^1(\bar{M})$ for all $k\in \mathbb{N}$ and hence the Sobolev embedding theorem \cite[Theorem 1.3.6]{S95} implies that $\omega\in \Omega^1(\bar{M})$. Lastly we can perform a Hodge-Morrey decomposition of $\omega=d\alpha+\delta\beta+\gamma$ for suitable smooth forms $\alpha,\beta,\gamma$ with $t(\alpha)=0=n(\beta)$ and $d\gamma=0=\delta\gamma$. Keeping in mind that $\omega$ may be approximated in $L^2$ by a sequence $\delta\Omega_k$ with $2$-forms $\Omega_k$ satisfying $n(\Omega_k)=0$, it is a direct consequence of Green's formula that $d\alpha=0=\gamma$ and hence in fact $\omega\in \Omega^1_n(\bar{M})$ as claimed. $\square$
\newline
\newline
\underline{Step 3: Characterisation of global minimisers}
\newline
\newline
Here we prove the characterisation of global minimisers as stated in \cref{T21}. The key is the following lemma
\begin{lem}
	\label{L43}
	Let $(\bar{M},g)$ be a $3$-manifold. Suppose $\alpha\in \Omega^1_n(\bar{M})$ is a Beltrami field corresponding to the eigenvalue $\lambda\in \mathbb{R}$, that is $\text{curl}(\alpha)=\lambda \alpha$. Then
	\begin{gather}
	\label{49}
	\mathcal{E}(\alpha)=\lambda \mathcal{H}(\alpha), \text{ where }\mathcal{E}(\alpha)=\left(\alpha,\alpha\right)_{L^2}.
	\end{gather}
\end{lem}
\underline{Proof of \cref{L43}:} The result follows immediately from Green's formula, keeping in mind the boundary conditions of elements of $\Omega^1_T(\bar{M})$. $\square$
\newline
\newline
If $h=0$ the characterisation is obvious. So without loss of generality we may assume that $h\neq 0$. We only consider the case $h>0$ because the arguments literally carry over to the case $h<0$, where simply some inequalities will be reversed due to the sign of $h$. We first claim that the eigenvalue $\lambda_{\omega}$ corresponding to any global minimiser $\omega\in \Omega^1_n(\bar{M})$ for $h>0$ is positive and in fact solely depends on the sign of $h$. First we fix some $h>0$ and assume that $\omega,\tilde{\omega}$ are both global minimisers within the same helicity class $h$. Then by the Beltrami field property and \cref{L43} we have
\[
\lambda_{\omega}=\frac{\mathcal{E}(\omega)}{\mathcal{H}(\omega)}=\frac{\mathcal{E}(\tilde{\omega})}{\mathcal{H}(\tilde{\omega})}=\lambda_{\tilde{\omega}},
\]
where we used that $\mathcal{H}(\omega)=\mathcal{H}(\tilde{\omega})$ because they both lie in the same helicity class and that their energies coincide because they are both global minimisers. This implies that the eigenvalue depends at most on the value of $h$ and not on any particular energy minimiser of a given helicity class. Let $0<h_1,h_2$ and suppose $\omega_1$, $\omega_2\in \Omega^1_n(\bar{M})$ are respective global energy minimisers with corresponding eigenvalues $\lambda_1,\lambda_2$. Define $\mu:=\sqrt{\frac{h_2}{h_1}}$ and $\hat{\omega}:=\lambda \omega_1$. Then we have $\mathcal{H}(\hat{\omega})=\lambda^2\mathcal{H}(\omega_1)=h_2$. In addition $\hat{\omega}$ is a multiple of $\omega_1$ and so it is also a Beltrami field corresponding to $\lambda_1$. We conclude from \cref{L43} and the fact that $\omega_2$ is a global minimiser within its helicity class
\[
\lambda_2\mathcal{H}(\omega_2)=\mathcal{E}(\omega_2)\leq \mathcal{E}(\hat{\omega})=\lambda_1\mathcal{H}(\hat{\omega})\Rightarrow \lambda_2\leq \lambda_1,
\]
where we used that the helicities coincide and are positive. By symmetry we obtain the reverse inequality, proving $\lambda_1=\lambda_2$. We denote this corresponding eigenvalue by $\lambda_+$. Since $\mathcal{E}(\omega_1)>0$ and $\mathcal{H}(\omega_1)=h_1>0$ we also immediately conclude from (\ref{49}) that $\lambda_+>0$. We claim that $\lambda_+$ is the smallest positive eigenvalue of curl$:\Omega^1_n(\bar{M})\rightarrow \Omega^1(\bar{M})$. To see this, suppose $\alpha\in \Omega^1_n(\bar{M})$ is an eigenfield corresponding to some eigenvalue $\lambda>0$. Since eigenfields are non-zero by definition we obtain from (\ref{49}) that $h:=\mathcal{H}(\alpha)>0$. By what we have shown so far we know that there exists some global minimiser $\omega\in \Omega^1_n(\bar{M})$ within the helicity class $h$ corresponding to the eigenvalue $\lambda_+$. Equation (\ref{49}) and the global minimiser property imply
\[
\lambda_+ h=\lambda_+ \mathcal{H}(\omega)=\mathcal{E}(\omega)\leq \mathcal{E}(\alpha)=\lambda \mathcal{H}(\alpha)=\lambda h\Rightarrow \lambda_+\leq \lambda,
\]
because $h>0$. This shows that $\lambda_+>0$ is indeed the smallest positive eigenvalue of the restricted curl operator as claimed. This proves the first implication of the characterisation of global minimisers.
\newline
For the converse implication let $h>0$ and $\omega\in \Omega^1_n(\bar{M})$ satisfy $\mathcal{H}(\omega)=h$ and $\text{curl}(\omega)=\lambda_+\omega$. We know that there exists a global minimiser $\hat{\omega}\in \Omega^1_n(\bar{M})$ within the same helicity class which corresponds to the eigenvalue $\lambda_+$. Then \cref{L43} implies
\[
\mathcal{E}(\omega)=\lambda_+\mathcal{H}(\omega)=\lambda_+h=\lambda_+\mathcal{H}(\hat{\omega})=\mathcal{E}(\hat{\omega}).
\]
Since $\hat{\omega}$ is a global energy minimiser, so must be $\omega$. $\square$
\newline
The inequalities in (\ref{22}) and (\ref{23}) also immediately follow from the considerations above.
\newline
\newline
\underline{Step 4: Real analyticity of local minimisers}
\newline
\newline
We will prove the following result, which obviously as a special case contains the desired interior regularity result
\begin{prop}
	\label{ZusatzProp}
	Let $(M,g)$ be an oriented, real analytic Riemannian $3$-manifold without boundary. If $X\in \mathcal{V}(M)$ is a smooth vector field and $\lambda, f\in C^{\omega}(M)$, i.e. real analytic functions into the real numbers, such that $\text{curl}(X)=\lambda X$ and $\text{div}(X)=f$, then $X$ is real analytic.
\end{prop}
\underline{Proof of \cref{ZusatzProp}:} Since writing out all the details will make the proof rather lengthy and since the main ingredients are standard elliptic estimates, we just give an outline of the proof here. According to \cite[Proposition 2.2.10]{KrPa02} it is enough to show that locally in some fixed open neighbourhood $U$ around any point $p$ we have: $\forall x_0\in U\text{ there is an} \text{ open neighbourhood }$ \\
$x_0\in V\subseteq U\text{ and }C,r>0: |\partial^{\alpha}X^j(x)|\leq C \frac{\alpha!}{r^{|\alpha|}}$ for every multi-index $\alpha\in \mathbb{N}^3_0$, every $1\leq j\leq 3$ and every $x\in V$. This will imply the real analyticity of $X$. In particular it is enough to locally control the $L^{\infty}$-norm of the derivatives of $X$ in a suitable manner.
\newline
To this end fix any $p\in M$ and let $\mu:U\rightarrow \mathbb{R}^3$ be a chart around $p$ with $\mu(p)=0$. For $0<R\ll 1$ suitably small we may equip $B_R:=B_R(0)\subset \mu(U)$ with the pullback metric $\tilde{g}:=\left(\mu^{-1}\right)^{\#}g$. Then the local expression $\tilde{X}:=\left(X^1\circ \mu^{-1},X^2\circ \mu^{-1},X^3\circ\mu^{-1}\right)$ of $X$ satisfies $\text{curl}_{\tilde{g}}(\tilde{X})=(\lambda\circ \mu^{-1})\tilde{X}$ and $\text{div}_{\tilde{g}}(\tilde{X})=f\circ \mu^{-1}$. Note that $\left(\overline{B_R},\tilde{g}\right)$ is a compact $\partial$-manifold in the sense of \cite{S95} and so $L^p$-norms induced by different metrics on $\overline{B_R}$ are equivalent. If we now let $Y$ be any smooth vector field on $\overline{B_R}$, which vanishes on the boundary, then this observation in combination with \cite[Lemma 2.4.10]{S95} implies an estimate of the form
\[
\sum_{i,j=1}^3\norm{\partial_iY^j}_{L^4(B_R)}\leq C\left(\norm{\text{curl}_{\tilde{g}}(Y)}_{L^4(B_R)}+\norm{\text{div}_{\tilde{g}}(Y)}_{L^4(B_R)} \right),
\]
where $L^4$ denotes the norm induced by the Euclidean metric and $C>0$ is some constant, not depending on $Y$. Now fix any $m\in \mathbb{N}$ and define for $0\leq k\leq m$, $\nu_k:=\frac{1}{3}\left(1+\frac{k}{m}\right)$. We can then for fixed $k$ choose a cutoff function $\eta\in C^{\infty}_c\left(B_{\nu_{k+1}R}\right)$ with the following properties
\[
0\leq \eta \leq 1,\text{ }\eta\equiv 1\text{ on }B_{\nu_kR}\text{ and }||\nabla \eta||_{\infty}\leq \frac{C}{3\nu_{k+1}-3\nu_k}=Cm\text{ for some }C>0.
\]
Here $C$ is independent of $m$ and from now on we will denote by $C$ generic constants, which may differ from line to line but are always independent of $m$ and $k$. Now fix any multi-index $\alpha\in \mathbb{N}_0^3$ and define $Y_{\alpha}:=\eta \partial^{\alpha}\tilde{X}$, where $\tilde{X}$ is the local expression of $X$. Applying our considerations so far to $Y_{\alpha}$ it is then standard to conclude that
\[
\sum_{i,j=1}^3\norm{\partial_i\partial^{\alpha}\tilde{X}^j}_{L^4(B_{\nu_kR})}\leq Cm\norm{\partial^{\alpha}\tilde{X}}_{L^4(B_{\nu_{k+1}R})}
\]
\[
+C\left(\norm{\text{curl}_{\tilde{g}}\left(\partial^{\alpha}\tilde{X}\right)}_{L^4(B_{\nu_{k+1}R})}+\norm{\text{div}_{\tilde{g}}\left(\partial^{\alpha}\tilde{X}\right)}_{L^4(B_{\nu_{k+1}R})} \right).
\]
Note that $\text{curl}_{\tilde{g}}$ does not commute with $\partial^{\alpha}$ since the metric explicitly depends on the point we consider it at. However keeping in mind the relations which $\tilde{X}$ satisfies and that we have a uniform bound on all derivatives, \cite[Proposition 2.2.10]{KrPa02}, of all real analytic quantities, we obtain an estimate of the form
\[
\sum_{i,j=1}^3\norm{\partial_i\partial^{\alpha}\tilde{X}^j}_{L^4(B_{\nu_kR})}\leq Cm \norm{\partial^{\alpha}\tilde{X}}_{L^4(B_{\nu_{k+1}R})}+CS\left(\alpha,\nu_{k+1}\right),
\] 
where $S\left(\alpha,\nu_{k+1}\right)$ is given by
\[
S\left(\alpha,\nu_{k+1}\right):= \frac{\alpha!}{r^{|\alpha|}}\left(\sum_{\beta\leq \alpha}\frac{r^{|\beta|}}{\beta!}\norm{\partial^{\beta}\tilde{X}}_{L^4(B_{\nu_{k+1}R})}+\sum_{\beta<\alpha}\frac{r^{|\beta|}}{\beta!}\sum_{i=1}^3\norm{\partial_i\partial^{\beta}\tilde{X}}_{L^4(B_{\nu_{k+1}R})}\right),
\]
for some suitable constant $r>0$ independent of $\alpha$, $m$ and $k$. Now if $|\alpha|=m$, we can iterate the above inequality $m$ times to obtain
\[
\norm{\partial^{\alpha}\tilde{X}}_{L^4(B_{\nu_0R})}\leq (Cm)^m\norm{\tilde{X}}_{L^4(B_{\nu_mR})}+C\sum_{k=1}^m(Cm)^{k-1}S\left(\alpha^{(k)},\nu_k\right)\text{, where }\nu_0=\frac{\nu_m}{2}=\frac{1}{3}
\]
and where $\alpha^{(1)}$ is obtained from $\alpha$ by subtracting $e_i$, the $i$-th standard basis vector, with $i$ being the minimal index for which $\alpha_i=\max_{1\leq j\leq 3}\alpha_j$ and $\alpha^{(k+1)}:=(\alpha^{(k)})^{(1)}$. From this it is possible to derive an estimate of the form
\[
\norm{\partial^{\alpha}\tilde{X}}_{L^4(B_{\frac{R}{3}})}\leq (|\alpha|C)^{|\alpha|}\left(\norm{\tilde{X}}_{L^4(B_{\frac{2R}{3}})}+1\right)\text{ for all }\alpha \in \mathbb{N}_0^3,
\]
where $C>0$ is a constant independent of $\alpha$. It then follows from Morrey's inequality and some further elementary estimates that there exist constants $b,c>0$, independent of $\alpha$, such that
\[
\norm{\partial^{\alpha}\tilde{X}}_{C^0(B_{\frac{R}{3}})}\leq b^{|\alpha|}\alpha!c\text{ for all }\alpha\in \mathbb{N}_0^3.
\]
As pointed out at the beginning of the proof this implies the claim. $\square$
\subsection{Proof of corollary 2.2}
Here we will prove that for given $X\in \mathcal{V}_n(\bar{M})$ and volume-preserving diffeomorphism $\psi:\bar{M}\rightarrow \bar{M}$, we have $\psi_{*}X\in \mathcal{V}_n(\bar{M})$, i.e. $\psi_{*}X$ has a well-defined helicity, and that the helicity is preserved under the action of volume-preserving diffeomorphisms. Both facts are straightforward observations which we formulate as lemmas. Then obviously \cref{T21} will imply \cref{C22}.
\begin{lem}
	\label{L51}
	Let $(\bar{M},g)$ be a $3$-manifold and let $\psi:\bar{M}\rightarrow \bar{M}$ be a volume-preserving diffeomorphism and $X\in \mathcal{V}_n(\bar{M})$. Then $\psi_{*}X\in \mathcal{V}_n(\bar{M})$.
\end{lem}
\underline{Proof of \cref{L51}:} We use the fact that if $\psi$ is a volume-preserving diffeomorphism and $Y$ any smooth vector field on $\bar{M}$, then we have
\begin{gather}
\label{51}
\omega^1_{\psi_{*}Y}=\star\left((\psi^{-1})^{\#}(\star \omega^1_Y) \right),
\end{gather}
where $f^{\#}$ denotes the pullback via a smooth function $f$. This formula can be easily proved keeping in mind the relation $\star\omega^1_Y=\iota_Y\omega_g$, where $\omega_g$ denotes the Riemannian volume-form and $\iota_Y$ denotes the contraction of a form with $Y$. From this it immediately follows that $\psi_{*}Y$ is divergence-free whenever $Y$ is divergence-free.
\newline
We now claim that if $v\in T_p\bar{M}$ for any fixed $p\in \partial\bar{M}$ is tangent to the boundary and $\phi:\bar{M}\rightarrow \bar{M}$ is any diffeomorphism, then $\phi_{*}v$ is also tangent to the boundary. To see this we observe that diffeomorphisms map the boundary to the boundary and hence $\phi(p)\in \partial \bar{M}$. Let $\mu:U\rightarrow \mathbb{H}^3$ be any fixed chart around $p$ into the upper half space. Since $v$ is tangent to the boundary we may express it locally as $v=\sum_{i=1}^2v^i\partial_i(p)$. Then picking $\mu\circ \phi^{-1}:\phi(U)\rightarrow \mathbb{H}^3$ as a chart around $\phi(p)$ we can express $\phi_{*}v$ locally, with respect to the chosen chart, as $\phi_{*}v=\sum_{i=1}^2v^i\hat{\partial}_i\left(\psi(p)\right)$, i.e. $\phi_{*}v$ is tangent to the boundary. Finally by definition we can write $\omega^1_X=\star d\omega^1_A$ for some smooth $1$-form $\omega^1_A$ with $t(\omega^1_A)=0$. We conclude by definition of $t$ and the preceding consideration that $t\left((\psi^{-1})^{\#}\omega^1_A \right)=0$ and that by (\ref{51}) $(\psi^{-1})^{\#}\omega^1_A$ defines a vector potential of $\psi_{*}X$. $\square$
\newline
\newline
The following lemma in the spirit of Arnold concludes the proof of the corollary
\begin{lem}
	Let $(\bar{M},g)$ be a $3$-manifold and let $\psi:\bar{M}\rightarrow \bar{M}$ be a volume-preserving diffeomorphism and $X\in \mathcal{V}_n(\bar{M})$. Then $\mathcal{H}\left(\psi_{*}X\right)=\mathcal{H}(X)$.
\end{lem}
\underline{Proof:} By the proof of the preceding lemma we know that if $\omega^1_A$ is a vector potential of $\omega^1_X$, then $(\psi^{-1})^{\#}\omega^1_A$ defines a vector potential of $\psi_{*}X$. Thus we may use this vector potential to compute the helicity. By properties of the $L^2$-inner product we have
\[
\mathcal{H}(\psi_{*}X)=\int_{\bar{M}}\left(\left(\psi^{-1}\right)^{\#}\omega^1_A\right)\wedge \star \omega^1_{\psi_{*}X}=\int_{\bar{M}}\left(\psi^{-1}\right)^{\#}\left(\omega^1_A\wedge \star \omega^1_X\right)=\int_{\bar{M}}\omega^1_A\wedge \star \omega^1_X=\mathcal{H}(X),
\]
where we used (\ref{51}). $\square$
\subsection{Proof of theorem 2.3}
We may adapt Arnold's reasoning from \cite{A74}, keeping in mind the Hodge-Morrey-Friedrichs decomposition for manifolds with boundary, to deduce the desired result. A little bit of care needs to be taken to deal with the boundary terms. We have seen during the proof of \cref{L51} that volume-preserving diffeomorphisms take divergence-free vector fields tangent to the boundary to the same type of vector fields. We state it as a separate lemma
\begin{lem}
	\label{L61}
	Let $(\bar{M},g)$ be a $3$-manifold. Let further $X\in \mathcal{V}_P(\bar{M})$ and let $\psi:\bar{M}\rightarrow \bar{M}$ be a volume-preserving diffeomorphism. Then $\psi_{*}X\in \mathcal{V}_P(\bar{M})$. Here $\mathcal{V}_P(\bar{M})$ denotes the set of all smooth vector fields on $\bar{M}$ which are divergence-free and tangent to the boundary of $\bar{M}$.
\end{lem}
\underline{Proof of \cref{T23}:}
\newline
In the spirit of Arnold, \cite{A74}, we fix any divergence-free vector field $Y$ which is tangent to the boundary. This vector field generates a global flow $(\phi_t)_t$, see also \cite[Proposition 1.1.8]{S95}. Now let $B_0\in \mathcal{V}_P(\bar{M})$ be any fixed vector field and let $\mathcal{V}_{B_0}(\bar{M})$ denote the set of all vector fields obtained from $B_0$ by the action of a volume preserving diffeomorphism. Assume that $B\in \mathcal{V}_{B_0}(\bar{M})$ solves the minimisation problem (\ref{27}). Obviously we have $(\phi_t)_{*}B\in \mathcal{V}_{B_0}(\bar{M})$ for all times $t\in \mathbb{R}$. Thus by properties of $B$ we must have $\mathcal{E}((\phi_0)_{*}B)=\mathcal{E}(B)\leq \mathcal{E}((\phi_t)_{*}B)$ for all times $t$. We may now define the following function
\[
f:\mathbb{R}\rightarrow \mathbb{R}, t\mapsto \mathcal{E}((\phi_t)_{*}B).
\]
Then $f$ has a global minimum at $t=0$ and we must have
\[
0=\frac{d}{dt}|_{t=0}f(t)=\int_{\bar{M}} \frac{d}{dt}|_{t=0}g\left((\phi_t)_{*}B,(\phi_t)_{*}B \right)\omega_g=2\int_{\bar{M}}g\left([B,Y],B\right)\omega_g=2\left(\omega^1_{[B,Y]},\omega^1_B\right)_{L^2},
\]
where $[\cdot,\cdot]$ denotes the Lie-bracket of vector fields. Using the identity $[B,Y]=\text{curl}(Y\times B)-Y\text{div}(B)+B\text{div}(Y)$, while keeping in mind that these vector fields are divergence-free, we obtain from Green's formula
\begin{gather}
\label{Zusatz}
0=\left(\star d\omega^1_{Y\times B},\omega^1_B\right)_{L^2}=\left(\omega^1_{Y\times B},\star d \omega^1_B \right)_{L^2}-\int_{\partial\bar{M}}\iota^{\#}\left(t(\omega^1_B)\wedge \star n(\star \omega^1_{Y\times B}) \right).
\end{gather}
We need now an additional argument to get rid of the boundary term and reproduce Arnold's result for the boundaryless case. To this end we note that $Y$ and $B$ are both tangent to the boundary, by choice of $Y$ and by \cref{L61}. So if we fix $p\in \partial\bar{M}$ either $B(p)$ and $Y(p)$ are linearly dependent, in which case we have $(Y\times B)(p)=0$ or they are linearly independent. In the latter case we can argue that $T_p\partial\bar{M}$ is two-dimensional and since $Y\times B$ is $g$-orthogonal to $Y$ and $B$, the vector $(Y\times B)(p)$ is normal to the boundary. By the duality relation we conclude $n(\star\omega^1_{Y\times B})(p)=\star t(\omega^1_{Y\times B})(p)=0$. Thus for all $p\in \partial\bar{M}$ we have $n(\star\omega^1_{Y\times B})=0$ and the boundary term vanishes. We conclude from (\ref{Zusatz})
\[
0=\langle Y\times B,\text{curl}(B) \rangle_{L^2}=\langle B\times \text{curl}(B),Y\rangle_{L^2},
\]
where we used the triple product rule. Since $Y$ was arbitrary it is a consequence of the Hodge-Morrey-Friedrichs decomposition, \cite{S95}, that $B\times \text{curl}(B)$ is a gradient field. $\square$
\subsection{Proof of proposition 2.4}
We recall that $\mathcal{V}_n(\bar{M})$ denotes the set of all smooth vector fields on $\bar{M}$ which admit a smooth vector potential which is normal to the boundary and that $\mathcal{V}_P(\bar{M})$ denotes the set of all smooth vector fields on $\bar{M}$ which are divergence-free and tangent to the boundary. The proposition states that equality between these two sets holds if and only if the first de Rham cohomology of $\bar{M}$ vanishes.
\newline
\newline
\underline{Proof of \cref{P24}:} By \cite[Theorem 2.6.1]{S95} we have $H_{dR}^1(\bar{M})\cong \mathcal{H}^1_N(\bar{M})$, where $\mathcal{H}^1_N(\bar{M})=\{\gamma\in \Omega^1(\bar{M})|d\gamma=0=\delta\gamma\text{ and }n(\gamma)=0 \}$. Hence it suffices to show that equality holds if and only if $\mathcal{H}^1_N(\bar{M})=\{0\}$.
\newline
\underline{$\Rightarrow$:} Assume that $\mathcal{V}_n(\bar{M})=\mathcal{V}_P(\bar{M})$. Let $\gamma\in \mathcal{H}^1_N(\bar{M})$, then the vector field associated with $\gamma$ is an element of $\mathcal{V}_P(\bar{M})$. Our assumption implies that $\gamma$ can be written as $\gamma=\delta\Omega$ for some suitable $\Omega\in \Omega^2(\bar{M})$ with $n(\Omega)=0$. An application of Green's formula, keeping in mind the boundary condition $n(\Omega)=0$, yields $\gamma=0$. Hence $\mathcal{H}^1_N(\bar{M})=\{0\}$ as claimed.
\newline
\underline{$\Leftarrow$:} Assume that $\mathcal{H}^1_N(\bar{M})=\{0\}$ and let $X\in \mathcal{V}_P(\bar{M})$. We know that $\delta\omega^1_X=0$ and $n(\omega^1_X)=0$. We can use the Hodge-Morrey decomposition to write $\omega^1_X=d\alpha+\delta\beta+\gamma$ where $\alpha,\beta,\gamma$ are smooth forms of appropriate degree with $t(\alpha)=0=n(\beta)$ and $d\gamma=0=\delta\gamma$. It follows immediately that $d\alpha=0$. We also know that $n(\delta\beta)=0$ because $n(\beta)=0$, \cite[Proposition 1.2.6]{S95}. In addition by linearity of $n$ and  since $n(\omega^1_X)=0$, we find $0=n(\omega^1_X)=n(\delta\beta)+n(\gamma)=n(\gamma)$. Overall we see that $\gamma\in \mathcal{H}^1_N(\bar{M})$ and hence by assumption $\gamma=0$. Therefore the Hodge-Morrey decomposition of $\omega^1_X$ simplifies to $\omega^1_X=\delta\beta$ with $n(\beta)=0$ $\Rightarrow \omega^1_X\in \Omega^1_n(\bar{M})$. $\square$
\section*{Acknowledgements}
I would like to thank Christof Melcher and Heiko von der Mosel for discussions and bringing the subject to my attention. This work has been funded by the Deutsche Forschungsgemeinschaft (DFG, German Research Foundation) – Projektnummer 320021702/GRK2326 –  Energy, Entropy, and Dissipative Dynamics (EDDy).
\bibliographystyle{plain}
\bibliography{mybibfile}
\footnotesize
\begin{flushleft}
	RWTH Aachen University, Lehrstuhl I f\"ur Mathematik, Turmstra{\ss}e 46, D-52064 Aachen, Germany
	\newline
	\textit{E-mail address:} \href{mailto:gerner@eddy.rwth-aachen.de}{gerner@eddy.rwth-aachen.de}
\end{flushleft}
\end{document}